# Impact of Gender and Age on performing Search Tasks Online

Georg Singer[1], Ulrich Norbisrath[1], und Dirk Lewandowski[2]

Institute of Computer Science, University of Tartu, Estonia[1]
Department Information, Hamburg University of Applied Sciences, Germany[2]

**Abstract**

More and more people use the Internet to work on duties of their daily work routine. To find the right information online, Web search engines are the tools of their choice. Apart from finding facts, people use Web search engines to also execute rather complex and time consuming search tasks. So far search engines follow the one-for-all approach to serve its users and little is known about the impact of gender and age on people´s Web search behavior. In this article we present a study that examines (1) how female and male web users carry out simple and complex search tasks and what are the differences between the two user groups, and (2) how the age of the users impacts their search performance. The laboratory study was done with 56 ordinary people each carrying out 12 search tasks. Our findings confirm that age impacts behavior and search performance significantly, while gender influences were smaller than expected.

## 1 Introduction

Due to the unprecedented growth of the Internet and it steadily being integrated into our daily routines, more and more tasks are carried out online. People perform online banking, read the news, communicate online, blog and are also immersed in other activities that require learning and decision making. A whole lot of information needs arise from the latter. As the amount of information available on the Internet is growing on a daily basis, search engines have become the dominant tools to search for information. Yet, search engines are machines to find documents based on keywords. Hence when it comes to more complex search tasks, search engines do not support those needs so well (White 2009; Marchionini 2006).

Especially as the Internet has grown into a mainstream medium, its users now comprise all population groups, women, men, children and elderly, tech savvy people and also beginners. Little is known so far about the differences of the behavior of theses user groups, in particular as far as complex search tasks are concerned. In addition, most of the user studies are done with non-representative groups like college students, who naturally show a different behavior than ordinary Web users due to their educational background. We assume that especially when it comes to more complex tasks, like planning a family gathering or a holiday trip online, or also researching information to make an important decision like purchasing an apartment, less Web-savvy users might be struggling more than expected. In those situations, when they work on more complex search tasks, they are required to review large amounts of different documents and Web sites, explore all the aspects relevant for their information need and finally might need to synthesize the outcome of their search into a single document of reference (White 2008; Singer 2012a). We present a study with a systematically selected user sample comprising of 56 ordinary Web users with very diverse backgrounds, roughly representing our society in terms of age (from 18 to 59) and gender. The study was done in a laboratory environment. All users and their behavior were logged using the Search-Logger tool (Singer et. al. 2011). We put the focus of this study on (1) examining the differences between men and women carrying out simple and complex search tasks and (2) finding out more about how age impacts the search behavior.

The rest of this paper is structured as follows: First, we review the literature on gender and age impact on search, followed by literature on simple, complex, and exploratory search. Then, we state some research questions. After that, we describe our methods, followed by the results. These are discussed, and in the conclusions section, we sum up the outcomes and limitations of our research.

## 2    Literature Review

Research on examining gender differences when carrying out complex or exploratory search tasks does not seem to exist. Lorigo et al. (2006) have used eye tracking to examine how different classes of users evaluate search engine results pages and found significant behavioral differences between men and women. Jackson et al. (2001) have carried out a survey with 630 Anglo-American undergraduates to examine their Internet usage patterns and according gender differences. They found out that women were mainly using the Internet for communicating (e-mail), while men were mainly searching for information. Hupfer and Detlor (2006) have carried out a survey-based study with 379 respondents, mainly students, to amongst others examine gender differences in Web information seeking. They present a self-concept orientation model and note that significant gender differences exist in web searching. While women use the Internet for communication and are interested in finding medical information and information about government and politics, men seem to be more interested in hobby-related information and investment and purchasing information. Liu and Hang (2008) have done a survey with 203 completed copies at a University campus in China with people aged between 18 and 23. Their findings are that female readers prefer reading from paper to reading online, and that there are significant differences between what to read and sustained attention. Roy and Chi (2003) conducted a study with 14 eighth grade students, 7 boys and 7 girls. The study participants had to carry out search tasks and were observed by two observers. Their findings show that boys used different search strategies than girls. Boys spent more time to enter queries and to scan and filter the hits on search engine results pages (horizontal search). When girls on the other hand clicked on search results, they browsed more deeply and investigated entire web sites more carefully (vertical searchers).

Regarding age and its influence on search behavior, a couple of studies can be found, but fewer than expected. Meyer et al. (1997) investigated the impact of age and training on Web search activity. In their study with 13 older and 7 younger users (ages not mentioned), they were able to show that the main difference between older study participants and younger ones was that both groups could fulfill most of the tasks, but it took the older ones more steps. Morrell et al. (2000) conducted a survey (consisting of 550 adults) to examine Web usage patterns among middle aged (ages 40-59), young-old (ages 60-74) and old-old adults (ages 75-92).  They report distinct age and demographic differences in individuals who use the Web. According to their study middle-aged and older Web users show similar behavior. Kubeck et al. (1999) examined the differences between older and younger adults finding information on the Web in a naturalistic setting. Their sample consisted of 29 older (mean 70.6 years) and 30 younger (mean 21.8) people. They were able to show that both groups found answers of similar quality but the older users where significantly less efficient in the process of searching. Aula (2005) has given a set of search tasks to 10 older adults. She observed the study participants while they carried out the tasks. She discovered that they were quite successful, but they had some operational difficulties in understanding how the Web was structured, with understanding the terminology and basic operations like text editing.  She proposed an "elderly friendly" search user interface. Dickinson et al. (2007) present a prototype for a Web search system for older people without any Internet experience. They also carried out a small user study and asked the users to rate the system against currently available mainstream search tools. The study confirms that older people search differently and have different requirements regarding user interface and usability.

When it comes to defining complex search and complex search tasks (as opposed to simple search tasks) it needs to be stated that a clear and simple definition is still missing. While Gary Marchionini (2006) indirectly defines exploratory search, which can be seen as being part of complex search (Singer et al. 2012a), as searching in a way which is not supported by today's search engines, and mentions eight aspects that characterize exploratory search, his definition has its shortcomings. Defining the concept based on what today's search engines cannot do, makes the definition unstable, as Web search engines are also developing and supporting a broader range of tasks. Task complexity can be defined objectively, independent of a person carrying out the task (March 1958; Shaw 1971). Another way to define the complexity of a task is by looking at the cognitive demands that are required from the task doer (Campbell 1986). Campbell (1988) proposes a classification for task complexity that is derived from a combination of the following complexity impacting factors: presence of multiple paths to a desired end-state, presence of multiple desired end states, presence of conflicting interdependence and presence of uncertainty or probabilistic linkages. A complex search task is the corresponding search task that leads to a complex search activity. A (complex) search task is therefore the description in contrast to (complex) search itself, which is the interactive process (Singer et al. 2012a). Other approaches like Interactive information retrieval (IIR) (Robins 2000) examine the longer term search process and work on reducing the artificial distinction between user and system. The already longer existing IIR and Exploratory Search are quite close. Exploratory search in comparison to IIR seems to point out the discovery aspect in search more. Other authors like Singer et al. (2012a) have decomposed the time-consuming complex search process into the concepts aggregation, discovery, and synthesis.

In contrast to many of the studies carried out in the papers just presented, in this present study we do not use a convenience sample made up of a small number of students or a specific group only but work with a systematically selected number of people representing a cross-section of society in terms of age, gender, and profession. A sample of 56 study participants is in comparison to the studies mentioned a relatively large number of participants. The value of our study not only lies in the results themselves, but also in validating (and falsifying) findings from older studies, based on a more realistic user sample.

# 3   Research Questions

To guide our research, we formulated the following research questions:

RQ1: What is the difference between women and men carrying out simple search tasks?

RQ2: What is the difference between women and men carrying out complex search tasks?

RQ3: What is the influence of age on the search behavior for simple search tasks?

RQ4: What is the impact of age on the search behavior for complex search tasks?

# 4   Research Method

This paper is based on a body of data which was collected in the course of one large experiment in August 2011. Two other articles (Singer et al. 2012b; Singer et al. 2012c) have been submitted for review so far. The first of those articles (Singer et al. 2012b) covers different aspects and the results presented are based on a distinct subset of data. The following description of the research design is based on that main paper. A more detailed description about the experimental set-up and the complete list of tasks used in the experiment can be obtained from there.

| Age span | Gender | | |
|---|---|---|---|
| | Female | Male | Total |
| 18-24 | 5 | 4 | 9 |
| 25-34 | 9 | 7 | 16 |
| 35-44 | 7 | 8 | 15 |
| 45-54 | 8 | 8 | 16 |
| 55-59 | 3 | 1 | 4 |
| Total | 32 | 28 | 60 |

*Table 1 User Sample*

The experiment was conducted in a laboratory environment. The carefully selected user sample originally consisted of 60 volunteers, roughly representing a cross-section of society in terms of age and gender (see Table 1 for further information on the participants). The effective number of study participants providing data to our study was 56 (30 men, 26 women), as the data of 4 (2 females, 2 males) out of the 60 users was corrupt and could therefore not be used. While we are well aware that such a sample size is still far from being perfectly representative, it is relatively large for a lab-based user study, and due to the wide span of users, the results from our study will be widely valid and not limited to a certain user group only as in many studies mentioned in the related work section. The experiment was conducted in Germany, and the language of the search tasks was German. Participants were recruited in various ways (e.g., through advertising).

The assignment for the study participants consisted of 12 search tasks – 6 simple ones and 6 complex ones. The answers had to be available somewhere in public websites in German as of August 2011. The 6 complex tasks were exploratory search tasks of varying complexity as defined by Kules and Capra (2009). We set the sequence of tasks up in a way, so that users could alternatively solve simple and complex ones. The aim was to keep the participants interested, and to not dis-encourage participants through a sequence of complex search tasks which they may be unable to solve.

Some of the six simple tasks (indicated by S), and six complex tasks (indicated by C)[1] were:

- (S) Joseph Pulitzer (1847-1911) was a well-known journalist and publisher from the US. The Pulitzer price is also carrying his name. In which European country was Pulitzer born?
- (S) How hot can it be on average in July in Aachen/Germany?
- (C) Are there differences regarding the distribution of religious affiliations between Austria, Germany and Switzerland? Which ones?
- (C) What are the most important 5 points to consider if you want to plan a budget wedding?

At the beginning each study participant completed a demographic form. In the demographic form we asked the users to fill in their gender and age. In addition we asked them about their Internet usage behavior, how often they were using the Internet per week, how long per day and for what purposes they were going online.

We used the following set of standard measures in our analysis:

- Ranking: We created a ranking of the users according to their performance in the whole experiment. We ranked the users first by the number of correct answers and then, in case of users with the same number of correct answers, by answers with right elements.
- SERP time: The time (s) users spent on search engine results pages (SERPs).
- Read time: The time (s) users spent on reading Web pages (other than SERP).
- Task time: The time it took users to finish a task
- Number of opened tabs: Number of browser tabs users opened per search task
- Number of queries: Number of queries users entered into search systems per task
- Average query length: Average number of words queries consisted of

---

[1] The complete list of tasks can be obtained from Singer et al. (2012b).

- Number of pages visited: Number of Web pages users visited per task

We have run paired-sample t-tests (assuming unequal variances) to analyze the statistical significance of our results.

| Simple tasks | Ranking (simple only) | SERP time (sec) | Read time (sec) | Number of tabs opened | Task time (sec) | Number of queries | Query-length (words) | Number of pages |
|---|---|---|---|---|---|---|---|---|
| female (n=30) | 3.8±0.6 | 32±5 | 96±9 | 4.6±0.7 | 128±13 | 1.9±0.2 | 2.9±0.2 | 2.2±0.1 |
| male (n=24) | 3.5±0.6 | 33±6 | 117±15 | 4.9±0.7 | 150±19 | 2.1±0.3 | 2.9±0.4 | 2.8±0.3 |
| p-value | 0.74 | 0.85 | 0.23 | 0.73 | 0.33 | 0.45 | 0.87 | 0.05 |
| interpretation | n.s. | n.s. | n.s. | n.s. | n.s. | n.s. | n.s. | s |

*Table 2 Comparison women and men carrying out simple search tasks*

## 5  Results

In this section we present the results, which answer our research questions.

RQ1: **What is the difference between women and men carrying out simple search tasks?**
Table 2 outlines our findings when comparing female and male users carrying out 6 simple search tasks as described in the methods section. It is evident that both user groups show quite similar behavior in case of simple search tasks. Despite some mean values being slightly differing, like average ranking and reading time (but the differences still being not statistically significant indicated by "n.s."), the only value with a significant (indicated by "s" in the last row) difference (underlined by a sufficiently small p-value of 0.05) is the number of pages, which users visited during their searchers (2.2 for women vs. 2.8 for men). The sample size here is only 54 (as opposed to 56) as the data of two users was corrupt.

| Complex tasks | Ranking (complex only) | SERP time (sec) | Read time (sec) | Number of tabs opened | Task time (sec) | Number of queries | Query-length (words) | Number of pages |
|---|---|---|---|---|---|---|---|---|
| female (n=30) | 5.6±0.5 | 101±10 | 320±30 | 3.8±0.4 | 421±32 | 5.4±0.4 | 4.0±0.3 | 6.8±0.7 |
| male (n=24) | 4.5±0.6 | 145±19 | 292±26 | 2.6±0.3 | 436±41 | 7.7±1 | 4.7±0.8 | 8.3±0.9 |
| p-value | 0.16 | 0.05 | 0.48 | 0.02 | 0.77 | 0.05 | 0.39 | 0.22 |
| interpretation | n.s. | s | n.s. | s | n.s. | s | n.s. | n.s. |

*Table 3 Comparison women and men carrying out complex search tasks*

RQ2: **What is the difference between women and men carrying out complex search tasks?** Table 3 shows the results for juxtaposing female and male users carrying out 6 complex search tasks as described in the method section. In the case of complex search tasks, the difference in behavior between the two genders becomes bigger. The measures SERP time, number of browser tabs opened, and number of queries are significantly different (supported by sufficiently small p-values). In case of ranking the p-values are a bit too high to make their mean values significantly different. Yet the difference of the two measures also indicates that ranking might be significantly different in case of a bigger or less diverse user sample. Again the sample size is only 54 (as opposed to 56), as the data of two users was corrupt.

| Simple tasks | Ranking (simple only) | SERP time (sec) | Read time (sec) | Number of tabs opened | Task time (sec) | Number of queries | Query-length (words) | Number of pages |
|---|---|---|---|---|---|---|---|---|
| age 1. quartile (n=11, 18-26) | 2.7±0.6 | 19±4 | 64±6 | 5.2±1.3 | 83±7 | 2.0±0.3 | 2.7±0.2 | 2.1±0.2 |
| age 4. quartile (n=11, 49-59)) | 4.0±0.8 | 40±8 | 146±17 | 5.2±1.2 | 186±23 | 1.8±0.3 | 2.3±0.3 | 2.4±0.4 |
| p-value | 0.21 | 0.03 | <0.01 | 0.99 | <0.01 | 0.57 | 0.29 | 0.53 |
| interpretation | n.s. | s | s | n.s. | s | n.s. | n.s. | n.s. |

*Table 4 Comparison of two age groups (1st quartile and 4th quartile) for simple search tasks*

RQ3: **What is the influence of age on the search behavior for simple search tasks?**

We are interested in examining the relation between age and the behavior when carrying out simple search tasks. Table 4 shows our findings, comparing selected measures for younger users (the first quartile of the user sample, consisting of 11 users and ages ranging from 18-26) and older users (fourth quartile of the user sample, consisting of 11 users and the age ranging from 49-59) carrying out simple search tasks. The task time is significantly smaller for the lower age group in comparison to the higher age group (83 sec. vs. 186 sec.). Apart from task time, also SERP time (19 sec. vs. 40 sec.) and read time (64 sec. vs. 146 sec.) are significantly different.

RQ4: **What is the impact of age on the search behavior for complex search tasks?**

We are interested in getting insight, how age influences the search behavior when carrying out complex search tasks. Table 5 shows our findings, comparing selected measures for younger users and older users carrying out complex search tasks (same quartiles of users as described in previous research question RQ3). For complex search tasks, the ranking is significantly better for younger users than for older ones (2.4 vs. 5.5). Also reading time (229 sec. vs. 434 sec.) and task time (333 sec. vs. 555 sec) are significantly smaller for the younger group than for the older group. The rest of the measures are not significantly different.

| Complex tasks | Ranking (complex only) | SERP time (sec) | Read time (sec) | Number of tabs opened | Task time (sec) | Number of queries | Query-length (words) | Number of pages |
|---|---|---|---|---|---|---|---|---|
| age 1. quartile (n=11, 18-26) | 2.4±0.5 | 104±19 | 229±17 | 3.0±0.6 | 333±33 | 6.8±0.8 | 4.1±0.5 | 6.4±0.7 |
| age 4. quartile (n=11, 49-59)) | 5.5±0.7 | 121±17 | 434±58 | 4.2±0.6 | 555±60 | 4.9±0.6 | 3.6±0.3 | 9.0±1.5 |
| p-value | <0.01 | 0.51 | <0.01 | 0.19 | <0.01 | 0.09 | 0.40 | 0.13 |
| interpretation | s | n.s. | s | n.s. | s | n.s. | n.s. | n.s. |

*Table 5 Comparison of two age groups (1st quartile and 4th quartile) for complex search tasks*

# 6   Discussion

In this section we are discussing and interpreting the results of the previous sections and also try to point out interesting findings and ideas. As expected, the way how men and women search is not significantly different when carrying out simple search tasks. The only measure that showed a significant difference was number of pages visited, which is significantly higher for men. It appears that men were browsing more to find the right information. That all the other measures are this similar also seems to confirm, that the standard procedure of issuing the query and finding the right information quickly with search engines works well in case of simple tasks and eventual gender-based differences in search strategies and the like simply do not appear. In case of complex search tasks, the situation is a bit different. Men spent significantly more time on SERPs, open significantly fewer browser tabs and issued a significantly higher number of queries. The fact that only three out of nine measures are significantly different seems to confirm the findings from the simple search tasks. Men and women show quite similar behavior when carrying out search tasks with Web search engines. We assume though that the sometimes high variances are partly due to the composition of our user sample (consists of people with very diverse backgrounds, from housewife to university student). Taking a more homogeneous user sample would most probably decrease the standard errors and give clearer signals. It would for example be interesting to see the results from a comparison between female housewives and male housemen only. When examining the impact of age on the way ordinary Web users carry out simple search tasks, our results show that significant differences between younger and older Web users are related to SERP time, read time and task time. Younger users are significantly quicker in carrying out the tasks. One would expect the rankings in the experiment to be different. Although the mean values are nominally different, due to the high standard error their difference is not significant. So it can be interpreted that it takes older people longer to carry out simple tasks, but they finally manage to find the same information as their younger counterparts. In case of complex search tasks the picture is quite different. The ranking for younger users was significantly better than for older users. Also the read time and task time were significantly smaller. In comparison to the simple tasks, it seems that in case of complex tasks, it took older people longer to carry out the task and also the outcome was not as good as for younger study participants. Differences in search capabilities become more evident the more complex a search task is.

## 7   Conclusion and Limitations

We presented the results of a study examining gender and age differences for a user sample of 60 ordinary Web users carrying out a set of 6 simple and 6 complex search tasks. This study is insofar quite unique, as we could not find many studies of this size, which were done with ordinary Web users (as outlined in the related work section). Most experiments are usually done with university students for various reasons (but also availability of cheap study participants). Therefore we doubt that the findings of many of those studies have a very general validity (apart from being for university students). Within our user sample of ordinary users (and very diverse experiences with web search) we found that the way men and women carry out simple search tasks is quite similar. Yet during their searches men visited a significantly higher number of Web sites. In case of complex tasks, the differences how men and women search became bigger. Women spent significantly less time on SERPs, opened more browser tabs and issued fewer queries per task. These findings are in line with the results presented by Roy and Chi (2003). When it comes to age and its impact on search behavior, in case of simple tasks it took older users significantly longer to execute the tasks and their SERP and read times also were significantly longer. In case of complex search tasks, younger users performed significantly better according to their ranking, they spent less time on reading and checking Web pages; in addition they carried out the complex tasks in a significantly smaller time. Especially the fact that younger users search more quickly than older ones is in line with the findings of (Morrell et al. 2009; Kubeck 1999; Aula 2005). Of course we also paid a price for the broadness of our user sample. Due to the very diverse backgrounds (from university student to housewife), we were faced with quite high variances in our numbers. This again resulted in high standard errors of mean. Hence we hypothesize that experiments with more focused user samples (like a younger and older group of academics only) might produce more significant differences.

**Acknowledgment**

This research was partly supported by the Estonian Information Technology Foundation (EITSA) and the Tiger University program, as well as by Archimedes.